\documentclass[superscriptaddress,letterpaper,amssymb,preprint,amsmath,titlepage]{revtex4-1}

\usepackage[english]{babel}
\usepackage[utf8]{inputenc}
\usepackage{mathtools}
\usepackage{mathrsfs}
\usepackage{braket}
\usepackage{xcolor}
\usepackage[symbol]{footmisc}
\usepackage{amsmath}
\usepackage{ragged2e}
\usepackage{xcolor, soul}
\setstcolor{red}
\setulcolor{blue}

\usepackage[labelfont=bf]{caption}
\DeclareCaptionJustification{justified}{\justifying}
\usepackage{subcaption}

\addto\captionsenglish{}
\usepackage{scalerel}
\usepackage{tikz}
\usetikzlibrary{svg.path}
\definecolor{orcidlogocol}{HTML}{A6CE39}
\tikzset{
  orcidlogo/.pic={
    \fill[orcidlogocol] svg{M256,128c0,70.7-57.3,128-128,128C57.3,256,0,198.7,0,128C0,57.3,57.3,0,128,0C198.7,0,256,57.3,256,128z};
    \fill[white] svg{M86.3,186.2H70.9V79.1h15.4v48.4V186.2z}
                 svg{M108.9,79.1h41.6c39.6,0,57,28.3,57,53.6c0,27.5-21.5,53.6-56.8,53.6h-41.8V79.1z M124.3,172.4h24.5c34.9,0,42.9-26.5,42.9-39.7c0-21.5-13.7-39.7-43.7-39.7h-23.7V172.4z}
                 svg{M88.7,56.8c0,5.5-4.5,10.1-10.1,10.1c-5.6,0-10.1-4.6-10.1-10.1c0-5.6,4.5-10.1,10.1-10.1C84.2,46.7,88.7,51.3,88.7,56.8z};}}
\newcommand\orcidicon[1]{\href{https://orcid.org/#1}{\mbox{\scalerel*{
\begin{tikzpicture}[yscale=-1,transform shape]
\pic{orcidlogo};
\end{tikzpicture}
}{|}}}}
\usepackage{xr-hyper}
\usepackage{hyperref} 
\hypersetup{colorlinks,citecolor=red,urlcolor=blue,hypertexnames=true}
\usepackage[capitalise]{cleveref}
\crefname{section}{SI section}{SI sections}
\usepackage{filecontents}

\makeatletter
\newcommand*{\addFileDependency}[1]{
  \typeout{(#1)}
  \@addtofilelist{#1}
  \IfFileExists{#1}{}{\typeout{No file #1.}}
}
\makeatother

\newcommand*{\myexternaldocument}[1]{%
    \externaldocument[SI:]{#1}%
    \addFileDependency{#1.tex}%
    \addFileDependency{#1.aux}%
}

\myexternaldocument{Paper_SI}
\begin{document}

\title{Self-heating of cryogenic high-electron-mobility transistor amplifiers and the limits of microwave noise performance}

\author{Anthony J. Ardizzi \orcidicon{0000-0001-8667-1208}}
\affiliation{Division of Engineering and Applied Science, California Institute of Technology, Pasadena, CA, USA}

\author{Alexander Y. Choi \orcidicon{0000-0003-2006-168X}}
\affiliation{Division of Engineering and Applied Science, California Institute of Technology, Pasadena, CA, USA}

\author{Bekari Gabritchidze \orcidicon{0000-0001-6392-0523}}
\affiliation{Division of Physics, Mathematics, and Astronomy, California Institute of Technology, Pasadena, CA 91125, USA}
\affiliation{Department of Physics, University of Crete, GR-70 013, Heraklion, Greece}

\author{Jacob Kooi \orcidicon{0000-0002-6610-0384}}
\affiliation{NASA Jet Propulsion Laboratory, California Institute of Technology, Pasadena, CA 91109, USA}

\author{Kieran A. Cleary}
\affiliation{Division of Physics, Mathematics, and Astronomy, California Institute of Technology, Pasadena, CA 91125, USA}

\author{Anthony C. Readhead}
\affiliation{Division of Physics, Mathematics, and Astronomy, California Institute of Technology, Pasadena, CA 91125, USA}

\author{Austin J. Minnich \orcidicon{0000-0002-9671-9540}
\footnote[1]{Corresponding author: \href{mailto:aminnich@caltech.edu}{aminnich@caltech.edu}}}
\affiliation{Division of Engineering and Applied Science, California Institute of Technology, Pasadena, CA, USA}

\date{\today} 
\begin{abstract}
The fundamental limits of the microwave noise performance of high electron mobility transistors (HEMTs) are of scientific and practical interest for applications in radio astronomy and quantum computing. Self-heating at cryogenic temperatures has been reported to be a limiting mechanism for the noise, but cryogenic cooling strategies to mitigate it, for instance using liquid cryogens, have not been evaluated. Here, we report microwave noise measurements of a packaged two-stage amplifier with GaAs metamorphic HEMTs immersed in normal and superfluid $^4$He baths and in vacuum from  1.6 -- 80 K. We find that these liquid cryogens are unable to mitigate the thermal noise associated with self-heating. Considering this finding, we examine the implications for the lower bounds of cryogenic noise performance in HEMTs. Our analysis  supports  the general design principle for cryogenic HEMTs of maximizing gain at the lowest possible power.
\end{abstract}
\maketitle

\section{Introduction}
Microwave low-noise amplifiers (LNAs) based on III-V semiconductor high electron mobility transistor (HEMT) technology \cite{pospieszalski_extremely_2005,bautista_chapter_2008} are a key component of high precision measurements across diverse fields in science and engineering such as radio astronomy \cite{pospieszalski_extremely_2018,chiong_low-noise_2022}, deep-space communication \cite{bautista_cryogenic_2001}, and quantum computing \cite{chow_implementing_2014,hornibrook_cryogenic_2015,krantz_quantum_2019,cha_300-w_2020,bardin_microwaves_2021}. In these applications, LNAs serve as the first or second stage of amplification in the receiver chain, thereby making a decisive contribution to the noise floor of the entire measurement apparatus. Although marked improvements in noise performance have been achieved in recent decades \cite{bautista_cryogenic_2001,wadefalk_cryogenic_2003,schleeh_ultralow-power_2012,akgiray_noise_2013,varonen_75116-ghz_2013,cuadrado-calle_broadband_2017,cha_0314_2018,heinz_noise_2020}, the noise performance of HEMT LNAs remains a factor of 3 -- 5 larger than the quantum limit \cite{chiong_low-noise_2022,cha_inp_2020,schleeh_cryogenic_2013}.

The noise behavior of HEMT amplifiers is typically interpreted using the Pospieszalski model \cite{pospieszalski_modeling_1989}. In this model, noise generators are assigned to the gate resistance at the input and the drain conductance at the output, parameterized by noise temperatures  $T_\text{g}$ and $T_\text{d}$, respectively. The drain temperature $T_\text{d}$ lacks an accepted physical origin, with several theories having been proposed \cite{statz_noise_1974,pospieszalski_limits_2017,gonzalez_noise_1997,esho_theory_2022}, and it is typically taken as a fitting parameter. The gate temperature $T_\text{g}$ is assumed to be equal to the physical device temperature \cite{pospieszalski_extremely_2005,bautista_physical_2007}. In this interpretation, cryogenic cooling leads to improvements in the noise figure of the HEMT in part by decreasing the gate temperature and hence its thermal noise.

The monotonic decrease in noise figure with decreasing physical temperature has been observed to plateau below physical temperatures of 20 -- 40 K (see Fig. 10 from Ref.~\cite{duh_32-ghz_1989}, Fig. 1 from Ref.~\cite{schleeh_phonon_2015} and Fig. 2 from Ref.~\cite{mcculloch_dependence_2017}, for example). Recent numerical and experimental studies have attributed this plateau to heating of the gate caused by power dissipated in the active channel, referred to as self-heating \cite{schleeh_phonon_2015,choi_characterization_2021}. In more detail, optimal low-noise performance at cryogenic temperatures requires power dissipation on the order of milliwatts. At these temperatures, the observed thermal resistance from Schottky thermometry is consistent with that expected of phonon radiation for which the thermal resistance varies as $T^{-3}$ \cite{choi_alexander_youngsoo_investigation_2022}. Consequently, at physical temperatures $\lesssim 20$ K the rapid increase in thermal resistance with decreasing temperature leads to a plateau in the gate temperature, which produces a corresponding plateau in noise figure.

Mitigating the effect of self-heating by enhancing heat dissipation is desirable. However, cryogenic thermal management of the gate in modern devices with sub-micron gate lengths and a buried gate structure is challenging. Existing on-chip cooling methods \cite{muhonen_micrometre-scale_2012} must be evaluated for their capability to provide cooling while avoiding detrimental impact on device noise performance. An alternate approach that does not require any device modifications is to submerge the heated surface in superfluid $^4$He, a quantum fluid with the highest known thermal conductivity \cite{van_sciver_helium_2012}. Such an approach is routinely used for cryogenic thermal management of superconducting magnets \cite{beig_applications_2001} and is actively employed in high-energy physics experiments \cite{kittel_cooling_1998,baudouy_heat_2014,weisend_ii_twenty-three_2016}. However, the effectiveness of liquid cryogens to mitigate self-heating in HEMTs has not been experimentally evaluated.

Here, we present noise measurements of a packaged two-stage amplifier with GaAs metamorphic HEMTs immersed in normal (He I) and superfluid (He II) $^4$He baths as well as in vapor and vacuum environments up to 80 K at various biases. We find that the liquid cryogens are unable to mitigate self-heating owing to the thermal boundary resistance between the HEMT surface and the $^4$He bath. We extract the gate temperature using a small-signal model of the device and show that the trends with physical temperature are generally consistent with those predicted by a phonon radiation model, regardless of the presence of liquid cryogens. We use these observations to examine the lower bounds of noise performance in cryogenic HEMTs, accounting for the temperature and power dissipation dependence of the thermal noise at the input.

\section{Experiment}

\subsection{Overview of measurement apparatus}\label{sec:expmeth}
We measured the microwave noise temperature $T_\text{e}$ and gain $G$ of the device under test (DUT), a common-source two-stage packaged amplifier comprised of OMMIC D007IH GaAs metamorphic HEMTs \cite{heinz_noise_2020}, each with a 70 nm gate length and a 4 finger 200 $\mu$m width gate structure consisting of an InGaAs-InAlAs-InGaAs-InAlAs epitaxial stack on a semi-insulating GaAs substrate with each stage biased nominally identically, using the cold attenuator Y-factor method \cite{leffel_y_2021}. An input matching network (IMN) was employed to match the optimal transistor impedance to the 50 $\Omega$ impedance of our measurement system over a 4--5.5 GHz bandwidth (see SI section \textcolor{cyan}{S.1} and Ch. 5.1 of Ref.~\cite{akgiray_ahmed_halid_new_2013} for more details).

\begin{figure}[t]
    {\phantomsubcaption\label{fig:setupSchem}}
    {\phantomsubcaption\label{fig:setupRaw}}
    \centering
    \includegraphics[width=1.0\textwidth]{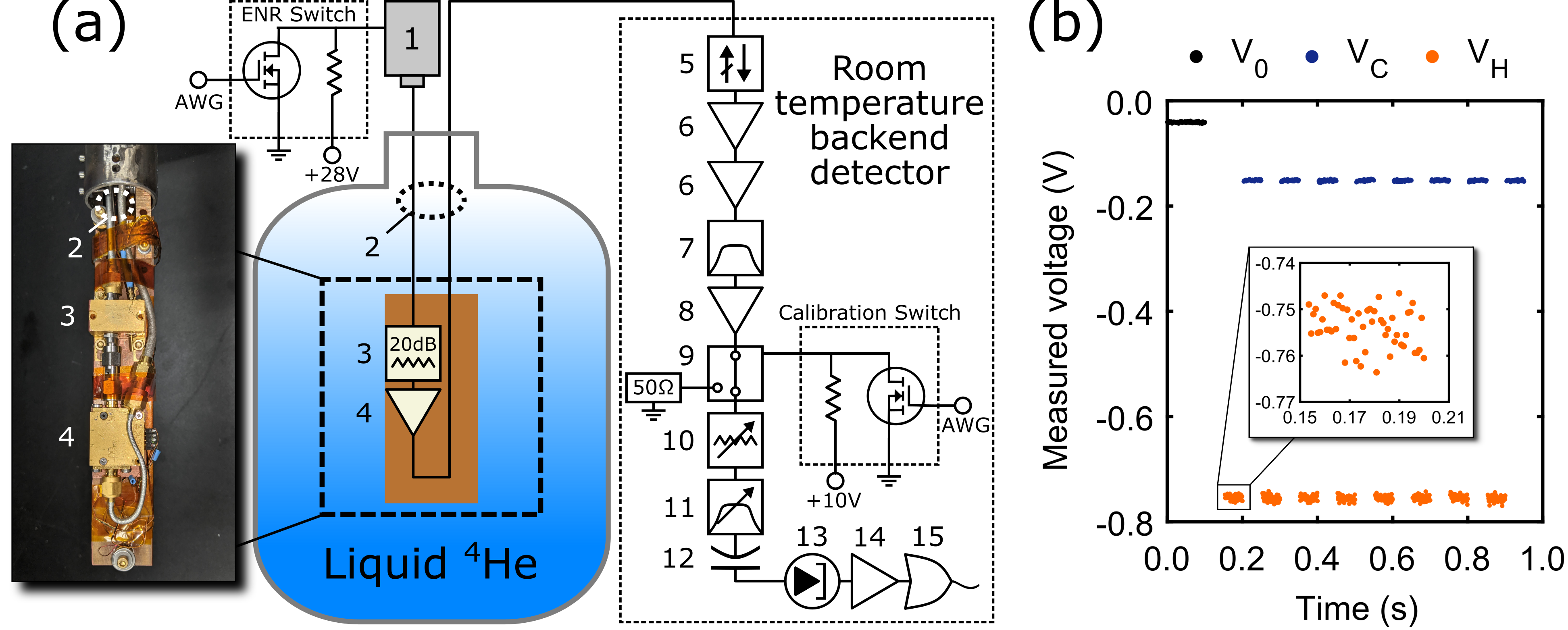}
    \caption{\textbf{(a)} Schematic of the microwave noise characterization apparatus designed for insertion into a $^4$He dewar. The image shows the copper mounting stage with the DUT, attenuator and temperature diodes mounted. A liquid $^4$He level sensor was mounted behind and above the copper stage inside the dipstick. The components consisted of (1) 15 dB ENR 2-18 GHz solid state SMA packaged noise diode biased at 28 V through a MOSFET biasing circuit, (2) input and output silver-plated stainless steel SMA coaxial cables each of length 1.3 m, (3) 20 dB packaged cryogenic chip attenuator with factory calibrated DT-670-SD diodes mounted directly on the attenuator substrate, (4) DUT, (5) Pasternak PE8327 isolator, (6) Minicircuits ZX60-83LN-S+ low-noise amplifiers, (7) Minicircuits filters with 3--6 GHz bandwidth, (8) Miteq AMF-3B-04000800-25-25P medium power amplifier, (9)  RF switch for calibration, (10) 0 -- 20 dB variable attenuator, (11) Micro Lambda MLFM-42008 20 MHz bandwidth tunable YIG filter, (12) Pasternak PE8224 inner-outer DC block, (13) Herotek DT4080 tunnel diode, (14) SRS560 low-noise preamp, (15) National Instruments NI6259-USB DAQ. The losses of the SMA cabling and attenuator pads are not shown.  \textbf{(b)} Representative raw noise power data versus time. The tunnel diode DC offset voltage $V_0$ (black symbols), hot voltage $V_\text{H}$ (orange symbols) and cold voltage $V_\text{C}$ (blue symbols) are all shown. The inset shows a zoom of one $V_\text{H}$ pulse.}
\end{figure}

\Cref{fig:setupSchem} shows a schematic of the measurement setup designed for microwave noise characterization in a liquid $^4$He dewar. The DUT, packaged 20 dB chip attenuator, temperature diodes and heater (mounted behind the stage and not shown) were screw-mounted to a copper mounting stage using indium foil. The stage was screw-mounted to a dipstick, and a liquid level sensor was taped to the interior wall of the dipstick. The dipstick was designed to mount on a 60 L dewar neck and submerge the stage in the liquid bath. A custom vacuum fitting with hermetic SubMiniature-A (SMA) and DC feedthroughs was sealed to the dewar neck, allowing for evaporative cooling via pumping. SMA coaxial cables and phosphor-bronze cryogenic wires were used to transmit microwave and DC signals, respectively, between the stage and feedthroughs. 

Noise power was generated by a packaged 2 -- 18 GHz solid state SMA noise diode with 15 dB excess noise ratio (ENR), which was biased using a MOSFET biasing circuit switched by an Agilent 33220A signal generator. This scheme allowed for Y-factor sampling up to 100 kHz, limited by the RC time constant of the MOSFET and noise diode circuit. Following the cold attenuator method \cite{fernandez_noise-temperature_1998}, this noise power was directed through 0.141'' diameter coaxial cabling to the 20 dB attenuator thermally strapped to the stage. The resulting hot and cold noise temperatures presented to the input plane of the DUT under cryogenic conditions were 52 K and 7 K respectively, yielding a Y-factor of 7.4. The noise power amplified by the DUT was then directed to a room temperature measurement apparatus. The $^4$He dewar was earthed, and inner-outer DC blocks were used to connect the hermetic feedthroughs to the noise source and backend to minimize low-frequency bias noise on the DUT.

\Cref{fig:setupSchem} also shows the room temperature backend measurement chain, which consisted of the following components. The microwave power emerging from the SMA feedthrough output first passed through a 3--6 GHz microwave isolator to improve impedance matching and minimize reflections. Two Minicircuits ZX60-83LN-S+ broadband amplifiers were placed sequentially following the isolator. A microwave filter was then used to limit the bandwidth of power amplified by the final gain stage, a Miteq AMF-3B-04000800-25-25P power amplifier. Next, a microwave switch was used to periodically switch between the signal path and a 50 $\Omega$ load, enabling re-calibration of the backend to correct for DC offset drifts. A variable attenuator and temperature-controlled YIG filter were then used to set the magnitude and frequency $f$ of microwave noise power reaching the Herotek DT4080 tunnel diode, which linearly transduced this power into a DC voltage. Inner-outer DC blocks were placed at the tunnel diode's input to eliminate unwanted biasing. The final DC signal was amplified and low-pass filtered by an SRS560 pre-amplifier and then digitized by an analog-digital converter for further data processing.

\Cref{fig:setupRaw} shows representative Y-factor voltage data spanning one duty cycle of data acquisition and backend re-calibration. Here, the hot (noise source on), cold (noise source off) and zero-power offset (50 $\Omega$ load) output voltages at time index $k$, denoted $V_{\text{H,}k}$, $V_{\text{C,}k}$ and $V_{0\text{,}k}$, respectively, were analog filtered at 1 kHz and over-sampled at $f_\text{s}=1.1$ kHz to avoid aliasing artifacts. The noise source was pulsed at $f_\text{ENR} = 10$ Hz, and the zero-power offset was re-calibrated at $f_0=1$ Hz with a 10\% duty cycle, yielding a voltage offset integration time of $t_0 = 0.1f_0^{-1}= 0.1$ s. Each half-pulse of hot (cold) voltage data was integrated for its duration $t_\text{ENR} = 0.5f_\text{ENR}^{-1}=0.05$ s. The first and last half-pulses of each cycle were discarded due to the $20$ ms switching time of the microwave switch, yielding 8 total pulses. The integrated zero-power offset was subtracted from each integrated hot (cold) half-pulse to give hot (cold) offset-corrected data $V_{\text{H},p}^\prime$ ($V_{\text{C},p}^\prime$) for each pulse $p$. This procedure yielded Y-factor data effectively sampled at $f_\text{ENR}$ with 0.2 s of data skipped every 1 s. The expression for the measured Y-factor for pulse $p$ considering the above specifications is:
\begin{equation}
    \label{eq:Yfac}
    Y_p = \frac{N_\text{HC}^{-1}\bigg(\sum\limits_{k=k_{\text{H},p}}^{k_{\text{HC},p}}V_{\text{H,}k}\bigg) - N_0^{-1}\bigg(\sum\limits_{k=1}^{N_0}V_{0\text{,}k}\bigg)}{N_\text{HC}^{-1}\bigg(\sum\limits_{k=k_{\text{HC},p}+1}^{k_{\text{C},p}}V_{\text{C,}k}\bigg) - N_0^{-1}\bigg(\sum\limits_{k=1}^{N_0}V_{0\text{,}k}\bigg)} = \frac{V_{\text{H},p}^\prime}{V_{\text{C},p}^\prime}
\end{equation}
where $p$ ranges from 1 to 8 for each cycle, $N_\text{HC}=t_\text{ENR}f_\text{s}$ is the number of sampled points in each half-pulse, $N_0=t_0f_\text{s}=2N_\text{HC}$ is the number of sampled points in each calibration pulse, $k_{\text{H},p}=(N_0+1)+2(p-1)N_\text{HC}$ is the time index at the start of the $p$th hot pulse, $k_\text{HC}=k_{\text{H},p}+N_\text{HC}$ is the time index at the center of the $p$th pulse when the noise source switches from hot to cold, and $k_{\text{C},p}=k_{\text{H},p}+2N_\text{HC}$ is the time index at the end of the $p$th cold pulse. For all steady-state data shown in this paper, the Y-factor was further averaged over a total measurement time $t_\text{fin} = 4$ s.

A Rhode\&Schwarz RSZVA50 VNA calibrated with a Maury 8050CK20 SOLT calibration kit was used for $S_{21}$ measurements. To measure gain and noise temperature simultaneously through Y-factor measurements, the gain $G_\text{full}=G L_\text{coax}^{-1}L_2^{-1}$ of the entire system from the noise source output plane to the backend input plane was first measured using the VNA, where $L_\text{coax}= L_1L_3$ is the total loss of the cables with input cable loss $L_1$, output cable loss $L_3$, and attenuator loss $L_2$. The DUT gain $G$ could then be extracted from the backend voltage under different conditions using:
\begin{equation}
    \label{eq:G}
    G = G_\text{full}L_\text{coax}L_2 = \frac{V_\text{H}^\prime-V_\text{C}^\prime}{V_\text{H}^\text{cal}-V_\text{C}^\text{cal}}G_\text{full}^\text{cal}L_\text{coax}L_2
\end{equation}
where $V_\text{H}^\text{cal}$, $V_\text{C}^\text{cal}$ and $G_\text{full}^\text{cal}$ are hot and cold output calibration voltages and total calibration gain (including cabling and attenuator), respectively, all measured at a particular device bias. The noise temperature $T_\text{e}$ was then determined by: 
\begin{equation}
    \label{eq:Te}
    T_\text{e} = \frac{1}{L_1L_2}\bigg[\frac{T_0E}{Y-1} - T_\text{C} - T_\text{coax}(L_1-1) - T_{L_2}(L_2-1)L_1 - \frac{T_\text{coax}(L_3-1)}{G_\text{full}L_3} - \frac{T_\text{BE}}{G_\text{full}}\bigg]
\end{equation}
where $T_0=290$ K, $E$ is the noise source excess noise ratio, $T_\text{C}$ is the noise source physical temperature, $T_\text{coax}$ is the lumped coaxial cable physical temperature, $T_2$ is the attenuator physical temperature, and $T_\text{BE}$ is the backend noise temperature (see SI section \textcolor{cyan}{S.2} for a derivation of \cref{eq:G,eq:Te}).

\subsection{Measurement apparatus calibration}\label{sec:calib}
We now describe the calibration procedure for each term of \cref{eq:G,eq:Te}. First, the backend noise temperature $T_\text{BE}$ was independently obtained using a liquid nitrogen cooled fixed load method. The noise source ENR was then calibrated by using the backend detector as a reference amplifier (see SI section \textcolor{cyan}{S.3} for more details). Two separate liquid $^4$He dewar baths were then used, one to calibrate the coaxial cable loss and temperature and then another to calibrate the attenuator loss.

The general procedure in each calibration dewar was as follows. The dipstick was used to submerge the mounting stage 2 cm from the bottom of a fully filled 60 L liquid $^4$He dewar at 4.2 K and ambient pressure. After waiting 30 minutes for thermal equilibration, calibration measurements were taken. The dewar was then sealed, and a Leybold DK50 rotary piston vacuum pump was used to evaporatively cool the liquid into the He II phase. An Anest Iwata ISP-500 scroll pump was connected in series and switched on after approximately 2 hours of pumping. A steady-state temperature of 1.6 K was reached (corresponding to a vapor pressure of 5.60 Torr) after roughly 6 hours of pumping, and further calibration measurements were taken as described in the following paragraphs. A heater was then switched on for less than 2 hours to accelerate the boil-off rate of the remaining liquid, and switched off when a spike in stage temperature was observed which indicated that the liquid surface had dropped below the stage. Further calibration was performed after turning off the heater as the stage was allowed to warm from 1.6 K under the ambient heating power of the measurement apparatus (see SI section \textcolor{cyan}{S.3} for more details). The dewar was then back-filled with $^4$He exchange gas to facilitate thermalization to room temperature, at which point the dipstick was removed. 
\begin{figure}[t]
    {\phantomsubcaption\label{fig:calibCoaxL}}
    {\phantomsubcaption\label{fig:calibCoaxT}}
    \centering
    \includegraphics[width=1.0\textwidth]{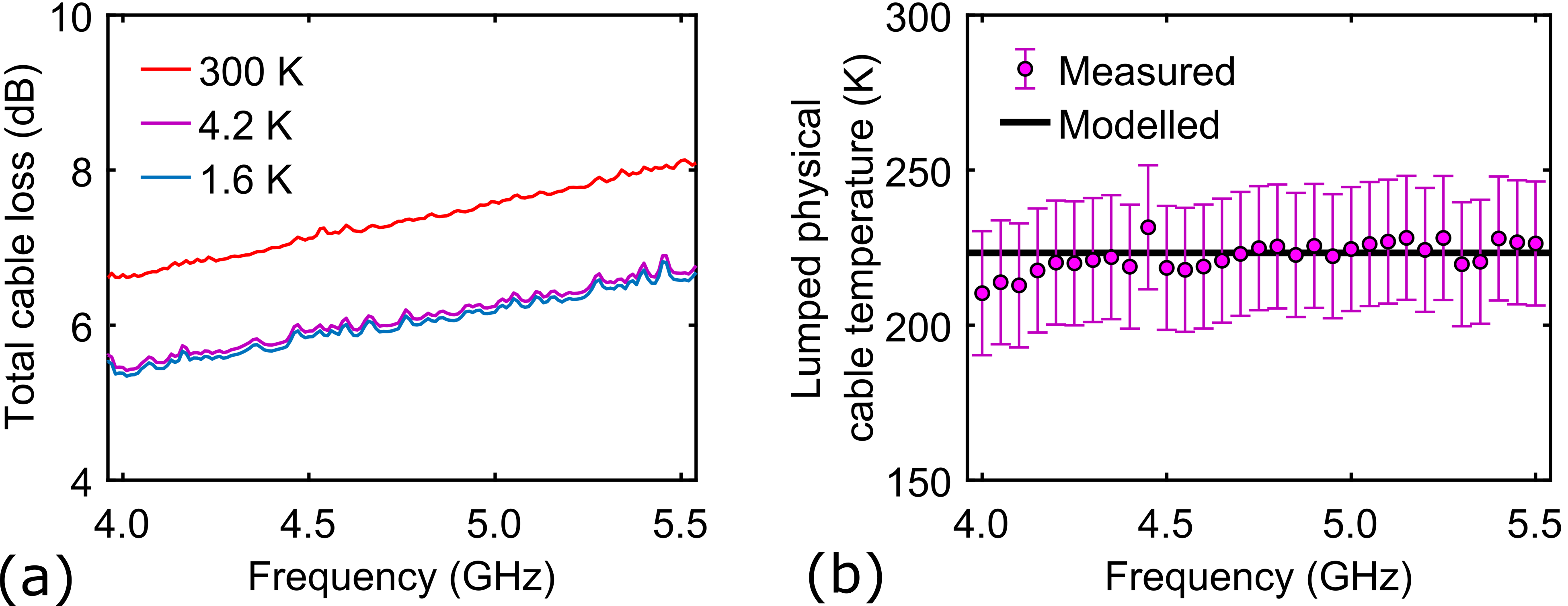}
    \caption{\textbf{(a)} Total loss of input and output coaxial cables versus frequency measured at 300 K (red line), 4.2 K (magenta line) and 1.6 K (blue line) with a commercial VNA. \textbf{(b)} Lumped physical coaxial cable temperature versus frequency obtained from Y-factor measurements (magenta circles) and from a heat conduction model (black line). Error bars represent an estimate of the total uncertainty including systematic errors.}
\end{figure}

We now discuss the details of the measurements in each calibration dewar. In the first calibration dewar, the attenuator and DUT shown in the configuration of \cref{fig:setupSchem} were replaced with a short through which was thermally anchored to the copper stage. \Cref{fig:calibCoaxL} shows the total loss $L_\text{coax}$ of the coaxial cables measured at room temperature in air and in the first calibration dewar at 4.2 K and 1.6 K. To isolate the individual cable losses $L_1$ and $L_3$, these losses were also measured independently at room temperature, and their ratio was assumed to remain constant for all temperatures.

The lumped coaxial cable physical temperature $T_\text{coax}$, which is the effective temperature at which the cables radiate their noise power, was measured directly using the Y-factor method. \Cref{fig:calibCoaxT} shows the measured lumped temperature versus frequency with the cables dipped in a 4.2 K He I bath. We assumed that both cables were at the same physical temperature. The 1.6 K He II calibration measurement is omitted for clarity as it is within 10 K of the He I measurement. To support the accuracy of the measured effective coaxial cable temperature, we estimated the value using an extension of the cable temperature model from Ref.~\cite{soliman_thermal_2016}. This estimate is shown in \cref{fig:calibCoaxT} and is in agreement with the measured values (see SI section \textcolor{cyan}{S.4} for more details).

In the second calibration dewar, the short through was replaced by the 20 dB attenuator. The total loss $L=L_\text{coax}L_2$ was measured using the same procedure as in the first calibration dewar, and the attenuator loss $L_2$ was extracted by dividing $L$ by the previously measured $L_\text{coax}$. A measurement of the temperature of the attenuator using the Y-factor method was not possible in this case due to the larger loss of the attenuator. Instead, the temperature $T_{L_2}$ of the attenuator was measured by a calibrated LakeShore DT-670-SD temperature diode indium-bonded directly to the attenuator chip. The diode calibration curve was provided by LakeShore, and the saturated liquid temperature of He I at 4.23 K was used to correct for DC offsets. We assumed that negligible temperature differences existed between the attenuator, mounting stage, and DUT, and we therefore took the DUT physical temperature to be $T_\text{phys} = T_{L_2}$. All temperature diodes were measured using a LakeShore 336 temperature controller, which converted the temperature to a voltage measured by the DAQ synchronously with all Y-factor measurements.

\subsection{LNA noise and gain measurements}
With the calibration data obtained, the noise temperature and gain of the DUT was measured using two additional liquid $^4$He dewar baths at various frequencies, temperatures, and biases. The gate-source and drain-source bias voltages $V_\text{GS}$ and $V_\text{DS}$, which were nominally applied equally to each individual transistor, were varied to yield transistor drain-source current densities $I_\text{DS}$ from 35 mA mm$^{-1}$ to 120 mA mm$^{-1}$, corresponding to dissipated power densities $P_\text{DC}$ per transistor of 15 mW mm$^{-1}$ to 150 mW mm$^{-1}$. The relevant measured values from the calibration procedure described above were used to extract $G$ and $T_\text{e}$ from \cref{eq:G,eq:Te}. 

In the first measurement dewar, the device was mounted to the stage in the configuration shown in \cref{fig:setupSchem}, and the cooldown procedure followed that of the calibration dewars described above. At each physical temperature, the DUT bias was varied and Y-factor measurements were taken versus frequency by adjusting the YIG filter frequency. In addition, the calibration measurement of $V_\text{H}^{\text{cal}}$, $V_\text{C}^{\text{cal}}$ and $G_\text{full}^{\text{cal}}$ was performed, with the DUT bias chosen for convenience to be its low-noise bias. Using these measurements, the device gain $G$ could be extracted from Y-factor voltage measurements at any bias using \cref{eq:G} without requiring a separate VNA measurement. This calibration was found to be stable over several days, and it was repeated each day before data acquisition.

In addition to measurements under specific liquid cryogen environments, continuous Y-factor measurements were also taken as the He II bath was pumped away, yielding noise data both before and after the DUT gate was submerged. The measurements were performed at a fixed bias of $I_\text{DS} = 80$ mA mm$^{-1}$ and frequency of $f=4.55$ GHz. The bias was chosen to be sufficiently high that any self-heating mitigation would be readily observed without risking device damage due to prolonged biasing, and the frequency was chosen due to it being the optimum noise match frequency as determined by the IMN \cite{akgiray_ahmed_halid_new_2013}. 

The He II film creep effect \cite{allen_properties_1939} was expected to cause the entire mounting stage, including the heated DUT region, to be coated in a superfluid film even after the liquid bath surface dropped below the DUT height. We used the sharp rise observed in the attenuator and stage temperature measurements to indicate the complete evaporation of He II from the stage. The DUT noise was also measured on warming from 1.6 K to 80 K in the vacuum space left after all liquid was pumped away. 
In the second measurement dewar, the noise temperature was again measured in 4.2 K liquid using the same procedure as in the first measurement dewar but without the subsequent evaporative cooling step. Instead, the liquid bath was allowed to evaporate under the heating power of the dipstick, enabling measurements to be taken in a vapor environment at several temperatures after the liquid level dropped below the mounting stage. The vapor warmed sufficiently slowly ($<$ 1 K/hour) such that all measurements were effectively taken in a steady state vapor environment. The calibrations used for these measurements were the same as for the 4.2 K liquid since the coaxial cable loss and temperature were observed to change negligibly up to 45 K stage temperature.

\section{Results}

\subsection{Microwave noise temperature versus frequency}
\begin{figure}[t]
    {\phantomsubcaption\label{fig:fswpA}}
    {\phantomsubcaption\label{fig:fswpB}}
    \centering
    \includegraphics[width=1.0\textwidth]{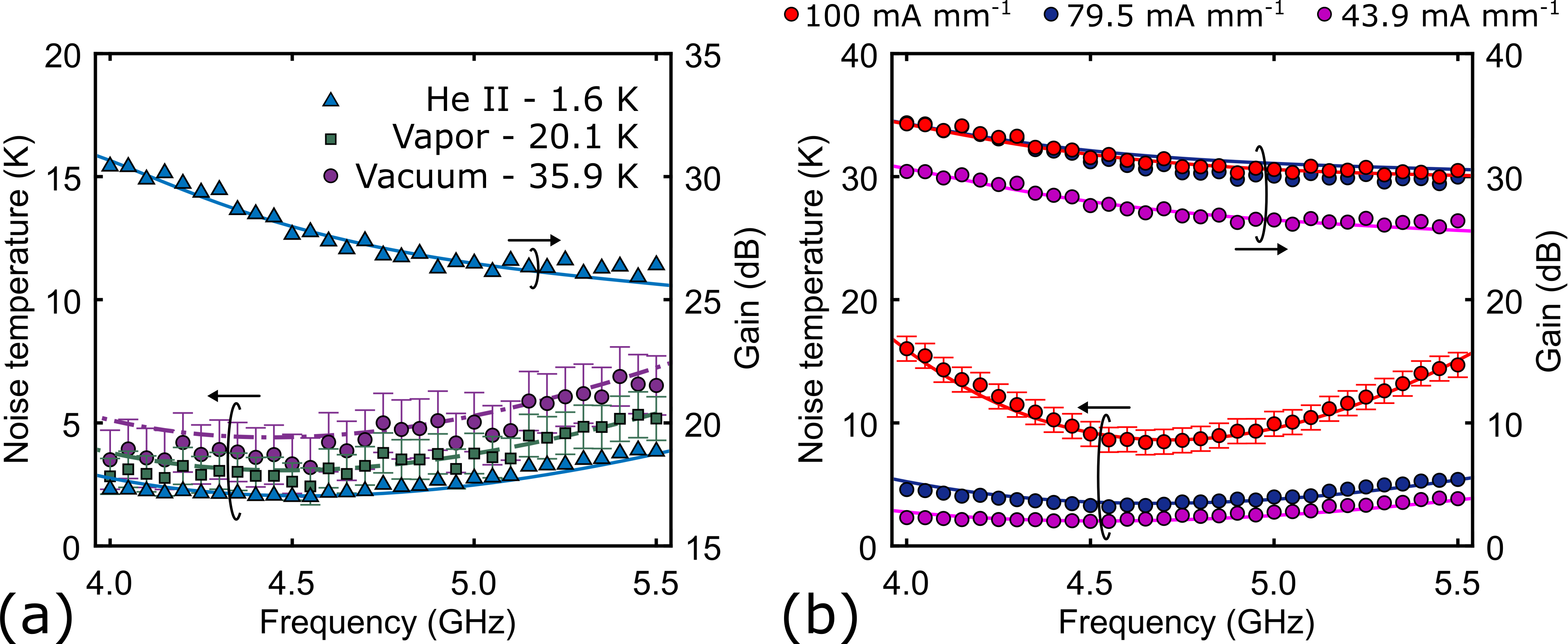}
    \caption{\textbf{(a)} Noise temperature (left axis) and gain (right axis) versus frequency, measured at the device's low-noise bias of $I_\text{DS}=43.9$ mA mm$^{-1}$ ($V_\text{DS}=0.56$ V, $P_\text{DC}=24.5$ mW mm$^{-1}$, $V_\text{GS}=-2.7$ V) in the following cryogenic environments: 1.6 K He II (blue triangles), 20.1 K vapor (green squares) and 35.9 K vacuum (purple circles). Only the gain under He II conditions is shown for clarity since the gain varies by less than 0.5 dB across all temperatures. The small-signal model fits for He II (solid blue line), $^4$He vapor (dashed green line) and vacuum (dash-dotted purple line) are also shown. \textbf{(b)} Noise temperature (left axis) and gain (right axis) versus frequency measured at biases of $I_\text{DS}=43.9$ mA mm$^{-1}$ (magenta circles;  $V_\text{DS}=0.56$ V, $P_\text{DC}=24.5$ mW mm$^{-1}$, $V_\text{GS}=-2.7$ V), $I_\text{DS}=79.5$ mA mm$^{-1}$ (dark blue circles; $V_\text{DS}=1.0$ V, $P_\text{DC}=79.5$ mW mm$^{-1}$, $V_\text{GS}=-2.7$ V) and $I_\text{DS}=100.0$ mA mm$^{-1}$ (red circles; $V_\text{DS}=1.2$ V, $P_\text{DC}=120$ mW mm$^{-1}$, $V_\text{GS}=-2.7$ V) with the DUT submerged in He II at 1.6 K. To vary the bias, the gate-source voltage was held constant at $V_\text{GS}=-2.7$ V while the drain-source voltage $V_\text{DS}$ was varied. The small-signal model fits (solid lines) are also shown. Where omitted in both \textbf{(a)} and \textbf{(b)}, the vertical error bars are equal to the height of the symbols.}
    \label{fig:fswp}
\end{figure}

We begin by showing the noise temperature and gain versus frequency at various temperatures, bath conditions and biases. \Cref{fig:fswpA} shows $T_\text{e}$ and $G$ versus $f$ with the device biased at its low-noise bias of $I_\text{DS}=43.9$ mA mm$^{-1}$ and immersed in three different bath conditions ranging from 1.6 K He II to 35.9 K vacuum, with measurements in 4.2 K He I and 8.2 K $^4$He vapor omitted for clarity since they are within within 0.2 K of the He II data. The noise increases monotonically with increasing physical temperature regardless of bath condition. At 4.5 GHz, the  noise temperature increases from 2.6 K to 3.3 K with temperature increasing from 20.1 K to 35.9 K, consistent with the expected $T_\text{e} \propto T_\text{phys}^{1/2}$ scaling predicted by the Pospieszalski model and observed in prior studies (see Fig. 2 from Ref.~\cite{pospieszalski_very_1994}, Fig. 1c from Ref.~\cite{schleeh_phonon_2015}, and Fig. 5 from Ref.~\cite{pospieszalski_dependence_2017}, for example). The gain varies by approximately 4 dB over the measured frequency range, peaking at $f=$ 4 GHz. The gain variation with physical temperature is less than 0.5 dB at fixed $f$ and $P_\text{DC}$, and so only the He II gain data is shown for clarity.

\Cref{fig:fswpB} shows $T_\text{e}$ and $G$ versus $f$ with the device immersed in 1.6 K He II at three different device biases of $I_\text{DS}=43.9$ mA mm$^{-1}$, $I_\text{DS}=79.5$ mA mm$^{-1}$, and $I_\text{DS}=100.0$ mA mm$^{-1}$. At biases below $I_\text{DS}=79.5$ mA mm$^{-1}$ the noise temperature varies by less than 1.5 K for all frequencies, whereas the noise temperature increases by 5.3 K from $I_\text{DS}=79.5$ mA mm$^{-1}$ to $I_\text{DS}=100$ mA mm$^{-1}$. The gain increases monotonically with increasing bias while retaining the same shape versus frequency, but it appears to asymptotically plateau at approximately the highest gain shown here, at a bias of $I_\text{DS}=100.0$ mA mm$^{-1}$.

To interpret these measurements, a microwave model of the full device including IMN, monolithic microwave integrated circuit (MMIC) components and transistor small-signal model was made using Microwave Office \cite{noauthor_awr_nodate}. The model used micrograph measured values of the IMN, foundry schematic values for the MMIC, and independently measured small-signal model values from nominally identical discrete transistors from Ref.~\cite{gabritchidze_experimental_2022}. The small-signal model and IMN parameters were manually tuned by less than 20\% from these starting values to fit both the gain and noise temperature curves (see SI section \textcolor{cyan}{S.1} for more details). Representative frequency-dependent results of the model are plotted in \cref{fig:fswpA,fig:fswpB}. The modelled and measured gain are in quantitative agreement over the full frequency range, and the model captures the overall trend in noise temperature as a function of temperature and bias.

Using this model, we  predict the expected change in device noise temperature in different cryogenic environments. Assuming $T_\text{g}$ changes from 20 K, the expected elevated gate temperature due to self-heating, to its lowest possible value of 1.6 K, and with all other small-signal parameters remaining unchanged, the model predicts that $T_\text{e}$ should decrease from 2.2 K to 0.6 K at $f=4.5$ GHz. As seen in both \cref{fig:fswpA,fig:fswpB}, the lowest measured noise temperature was $T_\text{e}=2.0\pm 0.2$ K, suggesting a gate temperature closer to 20 K. This initial finding suggests that the liquid cryogens are not altering the self-heating of the gate.

\begin{figure}[t]
    {\phantomsubcaption\label{fig:GatemodA}}
    {\phantomsubcaption\label{fig:GatemodB}}
    \centering
    \includegraphics[width=1\textwidth]{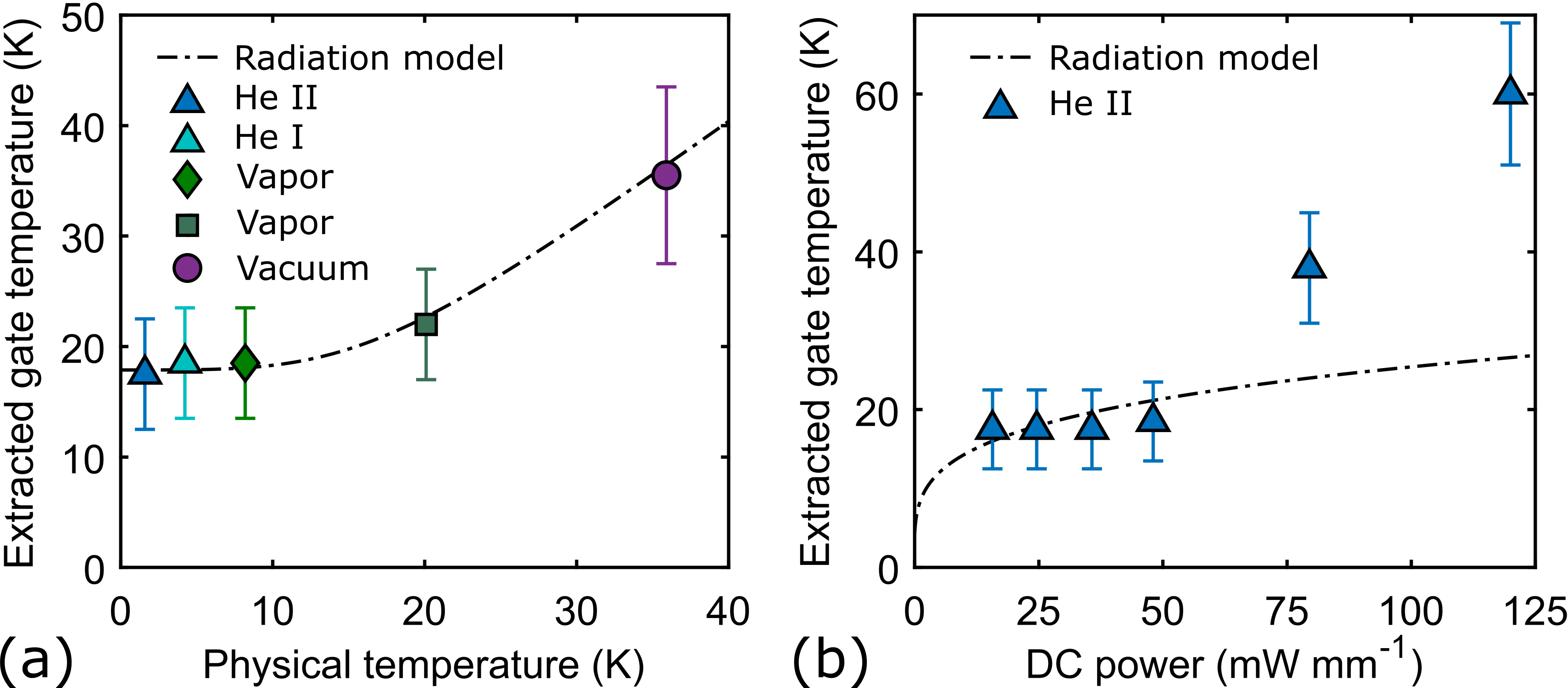}
    \caption{\textbf{(a)} Extracted gate temperature versus physical temperature at the device's low-noise bias of $I_\text{DS}=43.9$ mA mm$^{-1}$ ($V_\text{DS}=0.56$ V, $P_\text{DC}=24.5$ mW mm$^{-1}$, $V_\text{GS}=-2.7$ V). Symbols indicate extracted values and represent the same conditions as in \cref{fig:fswpA}, along with extracted values in 4.2 K He I (cyan triangles) and 8.1 K vapor (green diamonds). The radiation model is also shown (dash-dotted black line). \textbf{(b)} Extracted gate temperature versus bias power at 1.6 K physical temperature (blue triangles). The radiation model is also shown (dash-dotted black line). In both \textbf{(a)} and \textbf{(b)} the error bars were generated by determining the range of gate temperatures that accounted for the uncertainty in the frequency-dependent noise temperature data.}
    \label{fig:Gatemod}
\end{figure}

To obtain more quantitative insight, we used the model to extract the gate temperature $T_\text{g}$ under various conditions. \Cref{fig:GatemodA} shows the extracted $T_\text{g}$ versus $T_\text{phys}$ at the device's low-noise bias of $I_\text{DS}=43.9$ mA mm$^{-1}$. $T_\text{g}$ is elevated above $T_\text{phys}$ below 20 K even in the presence of superfluid, and equals $T_\text{phys}$ above 20 K, behavior which is in agreement with prior reports \cite{duh_32-ghz_1989,schleeh_phonon_2015,mcculloch_dependence_2017}. \Cref{fig:GatemodB} shows $T_\text{g}$ versus $P_\text{DC}$ at $T_\text{phys}=1.6$ K. Here, $T_\text{g}$ changes by less than 2 K for bias powers below $\sim50$ mW mm$^{-1}$, after which $T_\text{g}$ increases more rapidly.

We compare the small-signal model results with an equivalent circuit radiation model of the HEMT developed in Ref.~\cite{choi_characterization_2021}. The explicit functional form for the gate temperature derived from this model is:
\begin{equation}
    \label{eq:Tg}
    T_\text{g}(T_\text{s}, P_\text{DC}) = \bigg(T_\text{s}^4 + \frac{P_\text{DC}\mathscr{R}_{cs}\mathscr{R}_{gs}}{\sigma_p(\mathscr{R}_{cs}+\mathscr{R}_{gc}+\mathscr{R}_{gs})}\bigg)^{1/4}
\end{equation}
where $T_\text{s}$ is the substrate temperature, $\sigma_p=850$ W m$^{-2}$ K$^{-4}$ is the equivalent Stefan–Boltzmann constant for phonons in GaAs, and $\mathscr{R}_{ij}=A_{i}F_{ij}$ is the space resistance between nodes $i$ and $j$ with emitting line length $A_{i}$ and view factor $F_{ij}$ which quantifies the fraction of power emitted from surface $i$ that intercepts surface $j$. The subscripts $g$, $c$ and $s$ represent the gate, channel, and substrate, respectively. Following Ref.~\cite{choi_alexander_youngsoo_investigation_2022}, we take $A_\text{g}=A_\text{c}=70$ nm and compute $F_{gc}=0.3$.

The predictions of \cref{eq:Tg} are also shown in \cref{fig:GatemodA,fig:GatemodB}. In \cref{fig:GatemodA} the data and radiation model are in quantitative agreement over the full range of physical temperatures. We note that the model contains no fitting parameters. The extracted gate temperatures in the presence of liquid cryogens agree with the radiation model predictions. The increase in gate temperature with physical temperature above 20 K at fixed bias is also captured. In \cref{fig:GatemodB}, the data and model agree at biases below 50 mW mm$^{-1}$, but the data deviates from the model above $\sim50$ mW mm$^{-1}$. The origin of this discrepancy is presently unclear. A possible explanation is that other noise sources are being attributed to gate thermal noise, leading to artificially high extracted gate temperatures. This additional noise may signal the onset of impact ionization \cite{somerville_physical_2000,webster_impact_2000}, which is associated with a reduction in gain and an increase in gate leakage current. While the gain was observed to plateau at $33$ dB at the highest measured bias, as shown in \cref{fig:fswpB}, a relatively high gate leakage current of 100 $\mu$A was measured, suggesting some contribution of impact ionization. Excluding this non-ideal behavior at high biases, the good agreement in Fig.~\ref{fig:GatemodA} supports the phonon radiation mechanism of heat dissipation in cryogenic HEMTs, and it provides further indication that liquid cryogens provide inadequate  cooling power to mitigate self-heating in HEMTs.

\subsection{Noise temperature dependence on cryogenic environment}
\begin{figure}[t]
    {\phantomsubcaption\label{fig:TswpA}}
    {\phantomsubcaption\label{fig:TswpB}}
    \centering
    \includegraphics[width=1.0\textwidth]{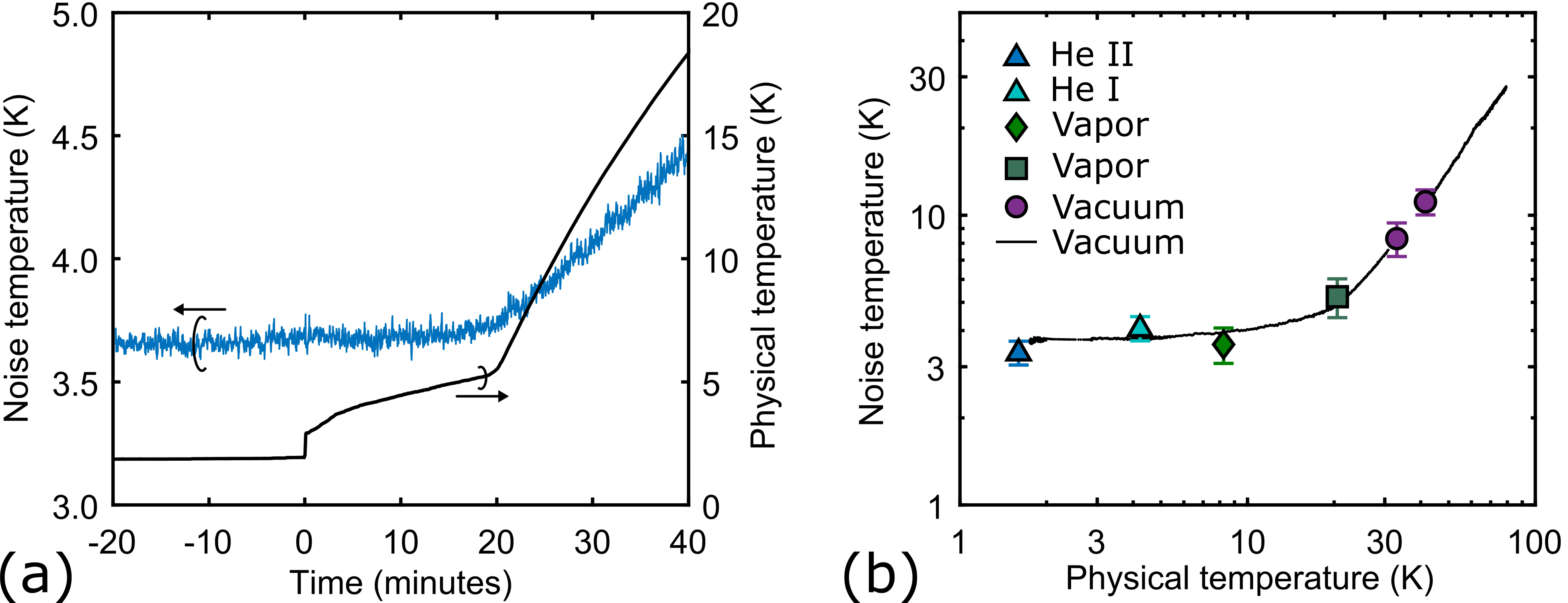}
    \caption{\textbf{(a)} Noise temperature (left axis, blue line) and physical temperature (right axis, black line) versus time in an evaporating He II bath sampled at $f_\text{ENR} = 10$ Hz and digitally filtered at 1 Hz, taken at a fixed bias $I_\text{DS} = 80$ mA mm$^{-1}$ ($V_\text{DS} = 1.0$ V, $P_\text{DC} = 80$ mW mm$^{-1}$, $V_\text{GS} = -2.8$ V) and frequency $f = 4.55$ GHz. The sharp kink in the physical temperature at time $t=0$ minutes, interpreted as the time at which superfluid is no longer present on the attenuator and device, is not reflected in the noise temperature. \textbf{(b)} Noise temperature (black line) versus physical temperature obtained from the transient data shown in \textbf{(a)}. Symbols show independently measured noise temperatures representing the same bath conditions as in \cref{fig:fswpA}, and the same bias conditions as in \textbf{(a)}. The presence of liquid cryogens does not affect the noise temperature within the measurement uncertainty.\label{fig:Tswp}}
\end{figure}

We obtain further insight into how liquid cryogens impact HEMT noise performance by examining the DUT noise temperature measured continuously in a changing cryogenic environment. \Cref{fig:TswpA} shows the time series of both $T_\text{e}$ and $T_\text{phys}$ measured continuously as the He II was pumped out of the measurement dewar. Here, $t=0$ minutes was chosen as a reference time at which a rise in $T_\text{phys}$ was observed, interpreted as the departure of superfluid from the attenuator and device. A corresponding feature in the DUT noise temperature is absent, suggesting that the superfluid cooling has no measureable effect on the DUT noise performance. After $t=0$ minutes the device thermalized with the surrounding $^4$He vapor, and $T_\text{e}$ was observed to increase smoothly with increasing physical temperature. After 20 minutes, the remaining He II liquid below the stage fully evaporated, and the warming rate increased as the mounting stage and DUT passively warmed to room temperature through the mounting apparatus. 

In \cref{fig:TswpB} the warming curve of $T_\text{e}$ plotted against $T_\text{phys}$ is shown from 1.6 K to 80 K, taken from the time series in \cref{fig:TswpA}. Again, the noise temperature measured in vacuum exhibits no sharp features, instead smoothly varying with physical temperature. Also plotted are noise temperatures measured separately under various bath conditions, at the same bias and frequency. The liquid, vapor and vacuum data all lie within the error bars of the warming curve. These observations suggest that the liquid and vapor cryogen environments provide no self-heating mitigation beyond maintaining a fixed ambient temperature. We note that the high-bias noise temperature exhibits a steeper slope with $T_\text{phys}$ than expected from the Pospieszalski model and that observed in \cref{fig:fswpA}. The origin of the discrepancy may be related to the contribution of impact ionization at this bias, but it does not affect the present discussion which depends only on the relative difference between noise temperatures in different cryogenic environments.

\section{Discussion}

\subsection{Limits on thermal conductance at the $^4$He-gate interface}
We consider our finding that He II is unable to mitigate self-heating in the context of prior studies of He II heat transport properties. We estimate the heat flow $\dot{Q}$ between the He II bath and the HEMT gate surface using $\dot{Q}=hA\Delta T$ where $h$ is the thermal boundary conductance, $A$ is the surface area of the gate, and $\Delta T$ is the steady-state temperature difference. We note that considerable uncertainty exists in all of these parameters owing to the complex heat transfer regime involving film boiling of superfluid $^4$He \cite{labuntzov_analysis_1979} and its dependence on the surface conditions, as well as the effective area for heat transfer of the HEMT. We therefore expect our estimate to give an order-of-magnitude indication of the heat flux.

To obtain the estimate, we take the surface area to be that of the gate head, 1 $\mu$m $\times$ 200 $\mu$m, which we take as an estimate of the heat transfer area of the dielectric passivation layer which covers the device.  A representative value of the temperature difference if the liquid cryogens measurably decreased the surface temperature is $\Delta T \sim 10$ K. To estimate the heat transfer coefficient, we take $h=6.2$ kW m$^{-2}$ K$^{-1}$, which is the highest measured He II thermal boundary conductance in the film boiling regime, taken from the compiled data in Table 7.5 from Ref.~\cite{van_sciver_helium_2012}. It is possible that $h$ could exceed this value owing to the micron-scale dimensions of the gate per Eq. 7.113 of \cite{van_sciver_helium_2012}, but we neglect this effect here. 

Assuming a temperature difference $\Delta T=10$ K, we estimate $\dot{Q} \sim 12$ $\mu$W. This value is two orders of magnitude less than the milliwatts of power required to optimally bias the device in our experiment. The absence of any observable effect of cooling by liquid cryogens in our experiment therefore indicates that the particular surface conditions of the HEMT do not increase the boiling heat flux sufficiently to make a measurable impact on the microwave noise temperature. This limitation might be mitigated to some by degree by, for instance, increasing the surface area with nanopatterned structures, so long as these structures do not impact noise performance; extensive further investigation would be required to evaluate this strategy.

\subsection{Implications for noise performance of cryogenic HEMTs}

We now examine the impact of self-heating on the noise performance of HEMTs assuming that thermal gate noise can only be reduced by decreasing the dissipated DC power. While the gate noise will indeed decrease with less power, the gain will also decrease, leading  to an increase in the contribution of both drain noise and any noise source originating after the gain stage of the HEMT, when referred to the input.

We first explore how $T_\text{min}$ from the Pospieszalski model \cite{pospieszalski_modeling_1989} varies with bias while including the explicit bias dependencies of both $T_\text{d}$ and $T_\text{g}$. In the limit $f \ll f_T$, $T_\text{min}$ is given by \cite{pospieszalski_extremely_2005}:
\begin{equation}
    \label{eq:Tmin}
    T_\text{min} = g_0\frac{\sqrt{T_\text{g}T_\text{d}}}{g_\text{m}} \\
\end{equation}
where we have explicitly introduced the prefactor $g_0 = 4\pi f(C_\text{gs}+C_\text{gd})\sqrt{(r_i+R_\text{G}+R_\text{S})g_\text{ds}}$ which we assume is bias-independent to isolate the bias dependence of $T\text{min}$ through $T_\text{g}$, $T_\text{d}$, and $g_0$. We assume $T_\text{phys}=4.2$ K, a gate-source capacitance $C_\text{gs}=150$ fF, a drain-source capacitance $C_\text{ds}=28$ fF, frequency $f=5$ GHz, a parasitic gate resistance $R_\text{G}=1\:\Omega$, a parasitic source resistance $R_\text{S}=1\:\Omega$, an intrinsic input resistance $r_\text{i}=1\:\Omega$, an intrinsic drain-source conductance $g_\text{ds}=15.4$ mS, and a transconductance $g_\text{m}$ obtained by taking a finite-difference approximation of the derivative of $I_\text{DS} - V_\text{DS}$ data for different $V_\text{GS}$ separated by 20 mV. All values are taken from Ref.~\cite{gabritchidze_experimental_2022}. We also assume a drain noise temperature that varies linearly from $T_\text{d}=20$ K at $I_\text{DS}=0$ mA mm$^{-1}$ to $T_\text{d}=1000$ K at $I_\text{DS}=100$ mA mm$^{-1}$, an approximation of the bias dependence measured in Ref.~\cite{gabritchidze_experimental_2022}, while taking $T_\text{d}=20$ K as the zero-bias limit.
\begin{figure}[t]
    {\phantomsubcaption\label{fig:PswpmodA}}
    {\phantomsubcaption\label{fig:PswpmodB}}
    \centering
    \includegraphics[width=1.0\textwidth]{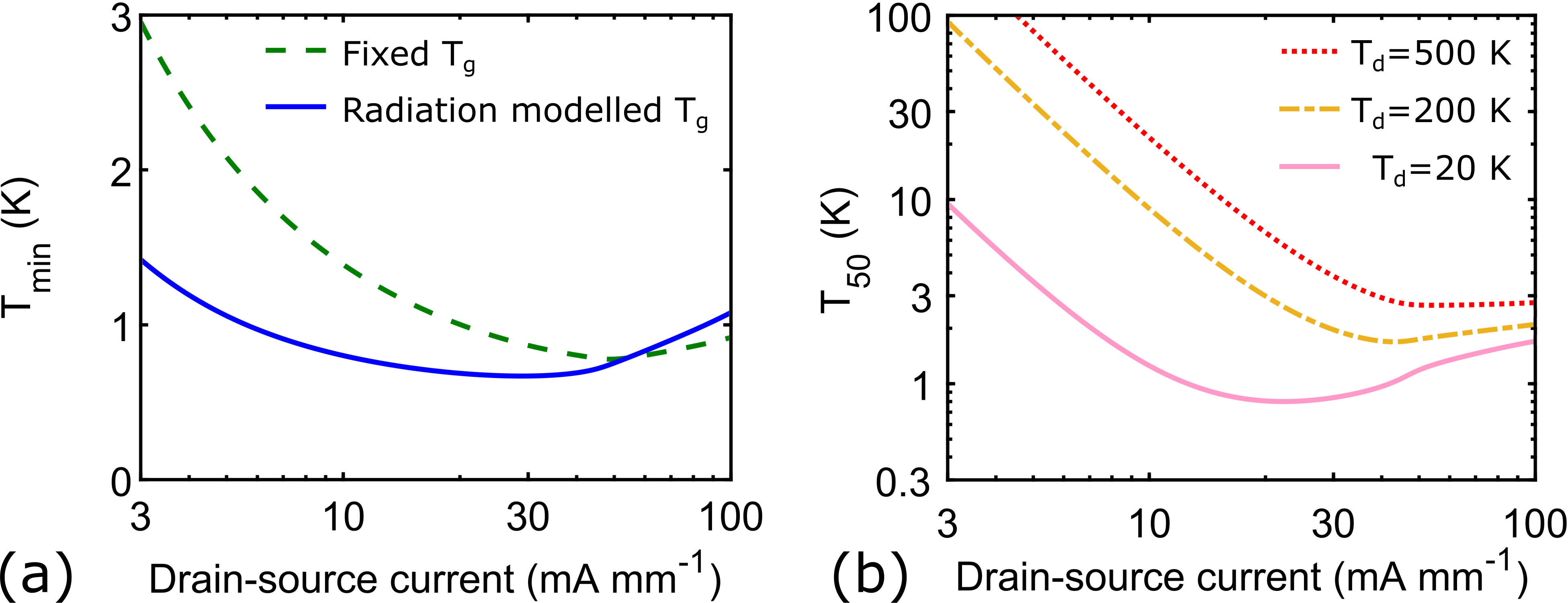}
    \caption{\textbf{(a)} Modelled $T_\text{min}$ versus drain-source current, shown for a fixed gate temperature $T_\text{g}=20$ K (dashed green line) and for a gate temperature with bias dependence determined by a radiation model (solid blue line). The radiation model predicts a gate temperature below $20$ K for powers below 40 mW mm$^{-1}$ (corresponding to $I_\text{DS}=54$ mA mm$^{-1}$), which is reflected in $T_\text{min}$. \textbf{(b)} Modelled $T_{50}$ noise temperature versus drain-source current, shown for $T_\text{d}=500$ K (dashed red line), $T_\text{d}=200$ K (dash-dotted gold line) and $T_\text{d}=20$ K (solid pink line). Both the minimum $T_{50}$ and the bias required to achieve this minimum decrease with decreasing $T_\text{d}$.}
    \label{fig:Pswpmod}
\end{figure}

\Cref{fig:PswpmodA} shows $T_\text{min}$ versus dissipated power both with and without the $T_\text{g}\propto P_\text{DC}^{1/4}$ dependence predicted by the radiation model. For the case of fixed gate temperature we assume $T_\text{g}= 20$ K. The radiation model predicts a lower $T_\text{min}$ than the fixed $T_\text{g}$ model up to a power of $P_\text{DC}=40$ mW mm$^{-1}$ (corresponding to $I_\text{DS}=54$ mA mm$^{-1}$), above which self-heating raises the gate temperature above 20 K. The minimum $T_\text{min}$ predicted by the radiation model is lower than that predicted by the fixed model by 0.12 K, and the bias at which the minimum $T_\text{min}$ occurs is also lower by 20 mA mm$^{-1}$. Below this optimal bias, both models predict an increase in $T_\text{min}$ with decreasing power, indicating where the gain is insufficient to overcome drain noise.

A reduction in drain noise at low biases is evidently beneficial for minimizing noise and power dissipation. We demonstrate the effect such a reduction has on the overall noise temperature by using a phenomenological model that both accounts for self-heating and separates the input and output noise contributions additively, a feature of noisy amplifiers which is not captured in the expression of $T_\text{min}$. At low enough frequencies such that $f \ll f_T$, the noise temperature of a HEMT with a 50 $\Omega$ source impedance is:
\begin{equation}
    \label{eq:T50}
    T_{50} = \frac{T_\text{g}(r_\text{i}+R_\text{G}+R_\text{S}) + T_\text{d}g_\text{ds}g^{-2}_\text{m}}{50 \: \Omega}
\end{equation}
as derived in Ref.~\cite{akgiray_ahmed_halid_new_2013} in the limit of open $C_\text{gs}$ and $C_\text{gd}$. For illustrative purposes, we assume a constant $T_\text{d}$, $T_\text{phys}=4.2$ K, and all other parameter values identical to those in the $T_\text{min}$ model.

\Cref{fig:PswpmodB} shows the modelled $T_{50}$ noise temperature versus dissipated power at different drain temperatures. Drain temperatures of 500 K and 200 K were chosen to approximate state-of-the-art low-power performance in GaAs devices \cite{gabritchidze_experimental_2022} and InP devices \cite{cha_300-w_2020}, respectively. A knee is observed in each curve, the location of which indicates where the gate noise and input-referred drain noise become comparable in magnitude. As $T_\text{d}$ is decreased, both the $T_{50}$ value at the knee and the bias at which the knee is observed decrease. This feature is explained as follows: as $T_\text{d}$ decreases, less gain is required to achieve the same contribution of $T_\text{d}$ to the overall noise, which implies that less power is required to bias the device, ultimately leading to less self-heating and therefore a lower $T_\text{g}$. In this way, reducing $T_\text{d}$ leads to a simultaneous improvement in noise temperature and reduction in optimal low-noise bias.

\section{Summary}
We have presented noise measurements of a packaged low-noise GaAs HEMT amplifier immersed in various cryogenic baths. The measured noise temperature and extracted gate temperature trends are generally consistent with those expected from heat dissipation by phonon radiation, independent of the presence of liquid cryogens. This finding indicates that self-heating of cryogenic GaAs metamorphic HEMTs cannot be mitigated, a result that is expected to extend to cryogenic HEMTs more generally. We explored the consequences of this result on overall noise performance for various values of the drain temperature. A decreased drain temperature was found to enable simultaneous improvements in noise performance by reducing the necessary gain and hence dissipated power, thereby reducing the self-heating. This result supports the general design principle of cryogenic HEMTs of maximizing gain at the lowest possible power.

\section{Supplementary Material}
The supplementary information provides further details on device modelling, a derivation of important equations from the main text, further calibration details, and error analysis.

\begin{acknowledgements}
The authors thank Sander Weinreb, Pekka Kangaslahti, Junjie Li, and Jan Grahn for useful discussions. A.A, A.Y.C.,  B.G., K.C., A.C.R., and A.J.M. were supported by the National
Science Foundation under Grant No. 1911220. Any opinions, findings, and conclusions or recommendations expressed in this material are those of the
authors and do not necessarily reflect the views of the National
Science Foundation. J.K. was supported by the Jet
Propulsion Laboratory PDRDF under Grant No. 107614-20AW0099.
Experimental work was performed at the Cahill Radio Astronomy
Laboratory (CRAL) and the Jet Propulsion Laboratory at the
California Institute of Technology, under a contract with the National
Aeronautics and Space Administration (Grant No. 80NM0018D0004).

\end{acknowledgements}

\bibliographystyle{aip}
\bibliography{references}

\end{document}


\title{Supplementary Information \\ Self-heating of cryogenic high-electron-mobility transistor amplifiers and the limits of microwave noise performance}

\author{Anthony J. Ardizzi \orcidicon{0000-0001-8667-1208}}
\affiliation{Division of Engineering and Applied Science, California Institute of Technology, Pasadena, CA, USA}

\author{Alexander Y. Choi \orcidicon{0000-0003-2006-168X}}
\affiliation{Division of Engineering and Applied Science, California Institute of Technology, Pasadena, CA, USA}

\author{Bekari Gabritchidze \orcidicon{0000-0001-6392-0523}}
\affiliation{Division of Physics, Mathematics, and Astronomy, California Institute of Technology, Pasadena, CA 91125, USA}
\affiliation{Department of Physics, University of Crete, GR-70 013, Heraklion, Greece}

\author{Jacob Kooi \orcidicon{0000-0002-6610-0384}}
\affiliation{NASA Jet Propulsion Laboratory, California Institute of Technology, Pasadena, CA 91109, USA}

\author{Kieran A. Cleary}
\affiliation{Division of Physics, Mathematics, and Astronomy, California Institute of Technology, Pasadena, CA 91125, USA}

\author{Anthony C. Readhead}
\affiliation{Division of Physics, Mathematics, and Astronomy, California Institute of Technology, Pasadena, CA 91125, USA}

\author{Austin J. Minnich \orcidicon{0000-0002-9671-9540}
\footnote[1]{Corresponding author: \href{mailto:aminnich@caltech.edu}{aminnich@caltech.edu}}}
\affiliation{Division of Engineering and Applied Science, California Institute of Technology, Pasadena, CA, USA}

\date{\today} 
\maketitle

\clearpage
\section{Device modelling}\label{sec:MWO}
\begin{figure}[ht]
    \centering
    \includegraphics[width=1.0\textwidth]{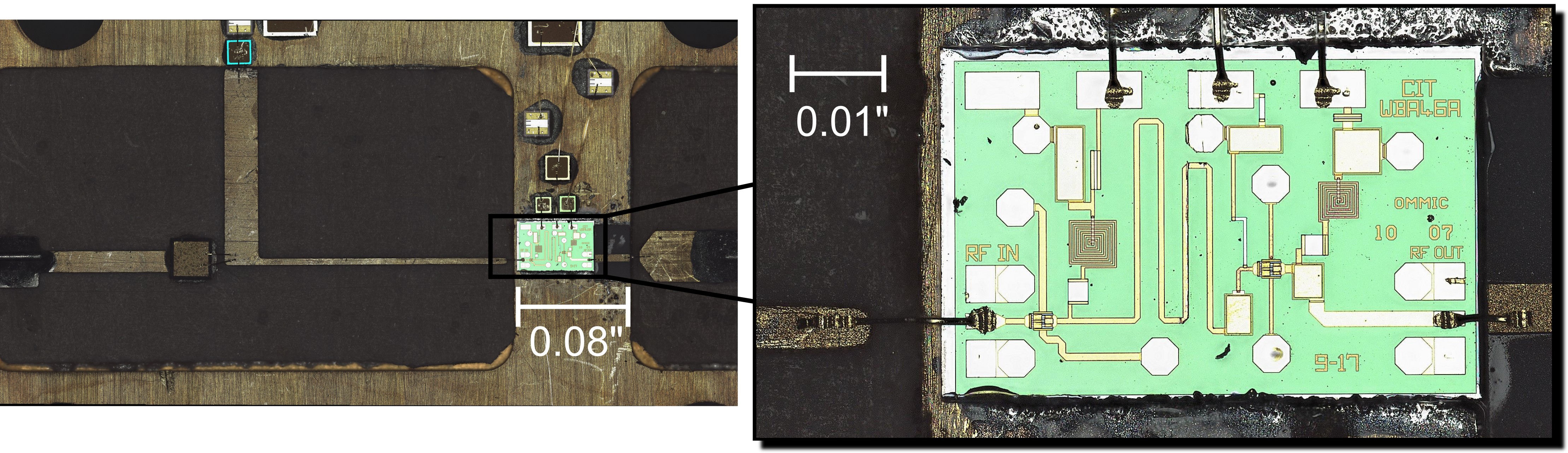}
    \caption{High-resolution micrograph image of the packaged device including the input matching network and MMIC. The inset shows a zoom of the MMIC fabricated by OMMIC.}
    \label{fig:MMICpic}
\end{figure}

Cadence AWR Microwave Office \cite{noauthor_awr_nodate} was used to create a model of the packaged device. \Cref{fig:MMICpic} shows a micrograph image of the 2-stage MMIC, including the external IMN, which are all housed inside a gold-plated copper chassis. All components of both the IMN and MMIC were included in the model. 

Model parameter fitting to the gain and noise temperature data was performed manually. The transistor small-signal model parameters were taken from nominally identical OMMIC devices measured in a separate study \cite{gabritchidze_experimental_2022}. All parameters were constrained to change by less then 20\% from these starting values. The gain data from several datasets were used first to tune the IMN microstrip geometry which determined the shape of the gain versus frequency. The gain data from each dataset was then used to tune the small-signal model parameters $C_\text{gs}$, $C_\text{gd}$, $g_\text{m}$ and $g_\text{ds}$. Finally the noise data was used to tune and extract $T_\text{g}$ and $T_\text{d}$. 

\clearpage
\section{Noise temperature and gain equation derivations}\label{sec:deriv}
We present a derivation of Eqs.\hspace{3pt}(\textcolor{cyan}{2}) and (\textcolor{cyan}{3}) from the main text, starting from the definition of the measured Y-factor: 
\begin{equation}
    \label{eq:Yfact}
    Y= \frac{P_\text{H}}{P_\text{C}} 
\end{equation}
where $P_\text{H}$ and $P_\text{C}$ are the noise powers presented to the input of the backend detector with the noise source turned on and off respectively. Using the definition of noise temperature $T_N=P_N(Bk_\text{B})^{-1}$ where $P_N$ is the noise power, $B$ is the measurement bandwidth and $k_\text{B}$ is the Boltzmann constant, we can write:

\begin{align}
    \frac{P_\text{H}}{Bk_\text{B}}&= \Bigg(T_\text{H}\frac{G}{L_1L_2L_3} + T_{L_1}\frac{G(L_1-1)}{L_1L_2L_3} + T_{L_2}\frac{G(L_2-1)}{L_2L_3} + T_e\frac{G}{L_3} + T_{L_3}\frac{(L_3-1)}{L_3} + T_\text{BE}\Bigg) \label{eq:PH} \\
     \frac{P_\text{C}}{Bk_\text{B}}&= \Bigg(T_\text{C}\frac{G}{L_1L_2L_3} + T_{L_1}\frac{G(L_1-1)}{L_1L_2L_3} + T_{L_2}\frac{G(L_2-1)}{L_2L_3} + T_e\frac{G}{L_3} + T_{L_3}\frac{(L_3-1)}{L_3} + T_\text{BE}\Bigg) \label{eq:PC}
\end{align}

where each term in \cref{eq:PH,eq:PC} represents the noise power added by successive elements in the measurement apparatus as defined in Section \textcolor{cyan}{IIA} of the main text, and $T_\text{H}=T_0E + T_\text{C}$ is the noise temperature of the noise source when switched on. We have also used the fact that the input-referred noise temperature $T_\text{in}$ of a matched attenuator with loss $L$ and physical temperature $T_L$ as derived in Ref. \cite{pettai_noise_1984} is: 
\begin{equation}
    \label{eq:Tatten}
    T_\text{in}=(L-1)T_L
\end{equation}
Plugging \cref{eq:PH,eq:PC} into \cref{eq:Yfact} and solving for $T_e$ yields Eq. (\textcolor{cyan}{3}) from the main text.

To derive Eq. (\textcolor{cyan}{2}) from the main text, we first recall that $P_\text{H}$ and $P_\text{C}$ are linearly transduced by the tunnel diodes with some power-to-voltage gain K so that we can write the measured voltages as $V_\text{H}= K P_\text{H}$ and $V_\text{C}= K P_\text{C}$ \cite{giordano_characterization_2016}. Plugging these into \cref{eq:PH,eq:PC}, subtracting \cref{eq:PH} from \cref{eq:PC}  and rearranging yields: 
\begin{equation}
    \label{eq:Glin}
    \frac{V_\text{H}-V_\text{C}}{G_\text{full}} = KBk_\text{B}T_0E
\end{equation}
Since the right side of this equation is constant, we see that the ratio $\frac{V_\text{H}^\prime-V_\text{C}^\prime}{G_\text{full}}$ must also be constant. This allowed us to calibrate the output voltage difference $V_\text{H}^\prime-V_\text{C}^\prime$ directly to the VNA measurement of the total gain $G_\text{full}$ by fixing the DUT bias to some arbitrarily chosen calibration value and measuring $V_\text{H}^\text{cal}$, $V_\text{C}^\text{cal}$ and the total gain $G_\text{full}^\text{cal}$ at this bias. The total gain at any other bias could then be extracted by measuring only $V_\text{H}^\prime-V_\text{C}^\prime$ and using \cref{eq:Glin}. Equation (\textcolor{cyan}{2}) from the main text follows from setting the left side of \cref{eq:Glin} equal to itself at two different DUT biases and rearranging for $G$.

\clearpage
\section{Calibration details}\label{sec:calib}
\begin{figure}[ht]
    \centering
    \includegraphics[width=1.0\textwidth]{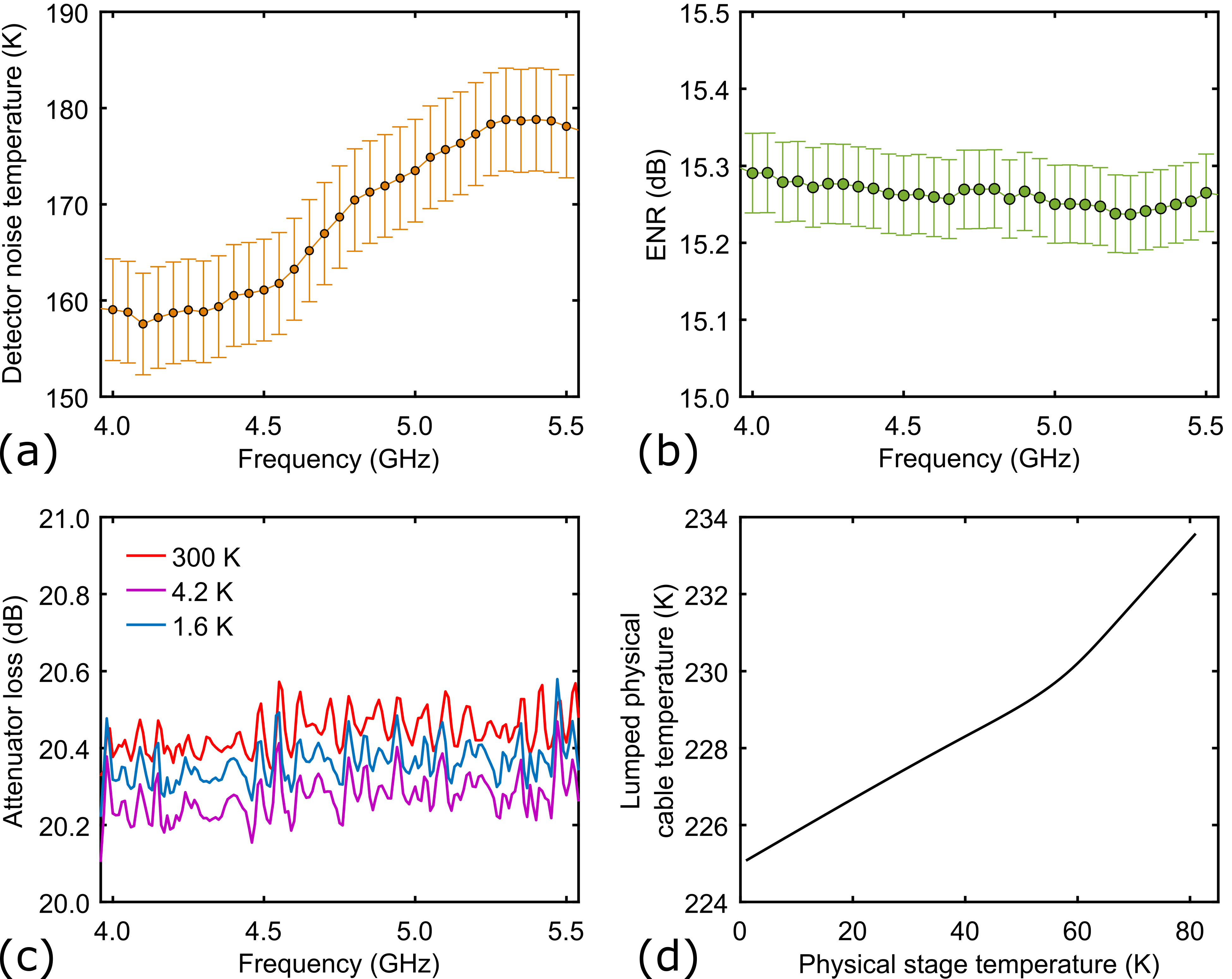}
    {\phantomsubcaption\label{fig:calibSI_T}}
    {\phantomsubcaption\label{fig:calibSI_ENR}}
    {\phantomsubcaption\label{fig:calibSI_Atten}}
    {\phantomsubcaption\label{fig:calibSI_Coax}}
    \caption{\textbf{(a)} Backend noise temperature versus frequency. Error bars reflect the uncertainty in the temperature of the cable connecting the 50 $\Omega$ load to the backend. \textbf{(b)} Noise source ENR versus frequency. Error bars reflect the error propagated from uncertainty in the backend noise temperature. \textbf{(c)} Attenuator loss versus frequency at room temperature (red line), 4.2 K (magenta line) and 1.6 K (blue line). \textbf{(d)} Lumped physical coaxial cable temperature versus physical stage temperature, taken from time series data of the cable temperature and stage temperature during the warming phase of the calibration measurements, and used as the calibration curve for the warming data shown in Fig. \textcolor{cyan}{5(b)} in the main text. 
    }
\end{figure}

\Cref{fig:calibSI_T} shows the calibration data for the backend noise temperature $T_{\text{BE}}$ versus frequency. This measurement was performed using a nitrogen cooled fixed load method, where the noise power of the backend terminated with a 50 $\Omega$ load was first measured at room temperature, and then measured again with the 50 $\Omega$ load submerged in a small liquid nitrogen bath at $77$ K. Thermal insulation was placed over the bath to minimize excessive cooling of the 3.5 inch long stainless steel coaxial cable connecting the backend to the 50 $\Omega$ load. The time between hot and cold measurements was approximately 1 minute, the time it took to dip the load into the nitrogen bath and allow its temperature to stabilize  as indicated by a plateauing of the measured noise power. All backend amplifiers, tunnel diodes and DC pre-amplifiers were wrapped in thermal insulation to mitigate drifts in gain and noise temperature.

\Cref{fig:calibSI_ENR} shows the calibration data for the noise source ENR versus frequency, measured by taking Y-factor measurements with the noise source connected directly to the calibrated backend detector. The ENR was extracted using the following equation:
\begin{equation}
    \label{eq:ENR}
    E = \frac{Y-1}{T_0}(T_\text{C}+T_\text{BE})
\end{equation}

The noise source chassis, which was wrapped in thermal insulation to promote thermal equilibration between the chassis and the internal noise diode, was monitored at all times using a type T thermocouple, and its temperature was found to vary negligibly under all experimental conditions. We took the chassis temperature to be equal to the internal diode temperature $T_\text{C}$.

\Cref{fig:calibSI_Atten} shows the attenuator loss versus frequency measured at room temperature, 4.2 K and 1.6 K. The measurement procedure is described in Section \textcolor{cyan}{IIB} of the main text. The losses varied by less than 0.1 dB between each temperature, which is consistent with other measurements of similar chip attenuators \cite{cano_ultra-wideband_2010}. 

\Cref{fig:calibSI_Coax} shows the lumped physical coaxial cable temperature versus stage temperature during the warming phase of the calibration measurements. Both of these quantities were measured independently as a time series, and the final calibration curve shown here was generated by fitting a smoothing spline to the raw data.

\clearpage
\section{Cable temperature calibration via the Y-factor method}\label{sec:cableYfac}
We measured the lumped input and output coaxial cable temperature $T_\text{coax}$ directly using the Y-factor method. Recalling \cref{eq:Tatten}, we can write the measured hot and cold powers from the Y-factor measurements of a single attenuator with loss $L$ and physical temperature $T_L$ as:

\begin{align}
    P_\text{H}&= \frac{T_0E+T_\text{C}}{L} + \frac{T_L(L-1)}{L} + T_\text{BE}\label{eq:PHatten}\\
    P_\text{C}&= \frac{T_\text{C}}{L} + \frac{T_L(L-1)}{L} + T_\text{BE}\label{eq:PCatten}
\end{align}
taking the ratio $Y=\frac{P_\text{H}}{P_\text{C}}$ and solving for $T_L$ gives:
\begin{equation}
    \label{eq:TL}
    T_L = \frac{1}{L-1} \Bigg(\frac{T_0E}{Y-1} - T_\text{C} - T_\text{BE}L \Bigg)
\end{equation}

To support the accuracy of this method, we developed a thermal model of the cables used in our experiment. This model extends the work from Ref. \cite{soliman_thermal_2016} to include the effect of heat transfer between the cables and the surrounding liquid and vapor. We consider infinitesimal cross-sectional slices of the cable in contact with the surrounding bath. Assuming steady-state where the net heat flux is zero we can write:
\begin{equation}
    \dot{Q}(x+\text{d}x) = \dot{Q}(x) + \text{d}\dot{Q}(x)
\end{equation}
where $\dot{Q}(x)$ is the heat flux at height $x$ along the cable, and $\text{d}\dot{Q}(x)$ is the differential convective heat flux along the slice of length $\text{d}x$. Applying Fourier's law along the length of the cable and using the definition of the convective heat transfer coefficient \cite{mills_heat_1999} we can write:
\begin{gather}
    \dot{Q}(x) = C_\text{s}(x)\frac{\text{d}T}{\text{d}x}\bigg|_x \\
    \text{d}\dot{Q}(x) = PH_\text{g,l}(x)\bigg(T(x)-T_\text{g,l}(x)\bigg)\text{d}x \\
    C_\text{s}(x) = \frac{\kappa_\text{steel}(x)A_\text{steel} + \kappa_\text{ptfe}(x)A_\text{ptfe}}{\kappa_\text{steel}(x)\kappa_\text{ptfe}(x)A_\text{steel}A_\text{ptfe}}
\end{gather}
where $C_\text{s}(x)$ is the thermal conductance arising from the parallel stainless steel and teflon heat conduction channels in the cable, $P$ is the perimeter of the cable, $H_\text{g,l}(x)$ and $T_\text{g,l}(x)$ are the convection coefficient and temperature of the surrounding gas and liquid baths, respectively, and $\kappa_\text{steel}(x)$, $\kappa_\text{ptfe}(x)$, $A_\text{steel}$ and $A_\text{ptfe}$ are the thermal conductivities and cross-sectional areas of the inner stainless steel conductor and teflon dielectric in the cable, respectively. The temperature-dependent thermal conductivities $\kappa_\text{steel}$ and $\kappa_\text{ptfe}$ were taken from compiled data from NIST \cite{noauthor_nist_nodate-1,noauthor_nist_nodate}. Massaging the above equations and rearranging, we arrive at:
\begin{equation}
    \label{eq:ThermProf}
    C_\text{s}(T(x))\frac{\text{d}^2T}{\text{d}x^2} + \frac{\text{d}C_\text{s}}{\text{d}T}(T(x))\bigg(\frac{\text{d}T}{\text{d}x}\bigg)^2 -PH_\text{g,l}(x)\bigg(T(x)-T_\text{g,l}(x)\bigg) = 0
\end{equation}

\begin{figure}[ht]
    \centering
    \includegraphics[width=1.0\textwidth]{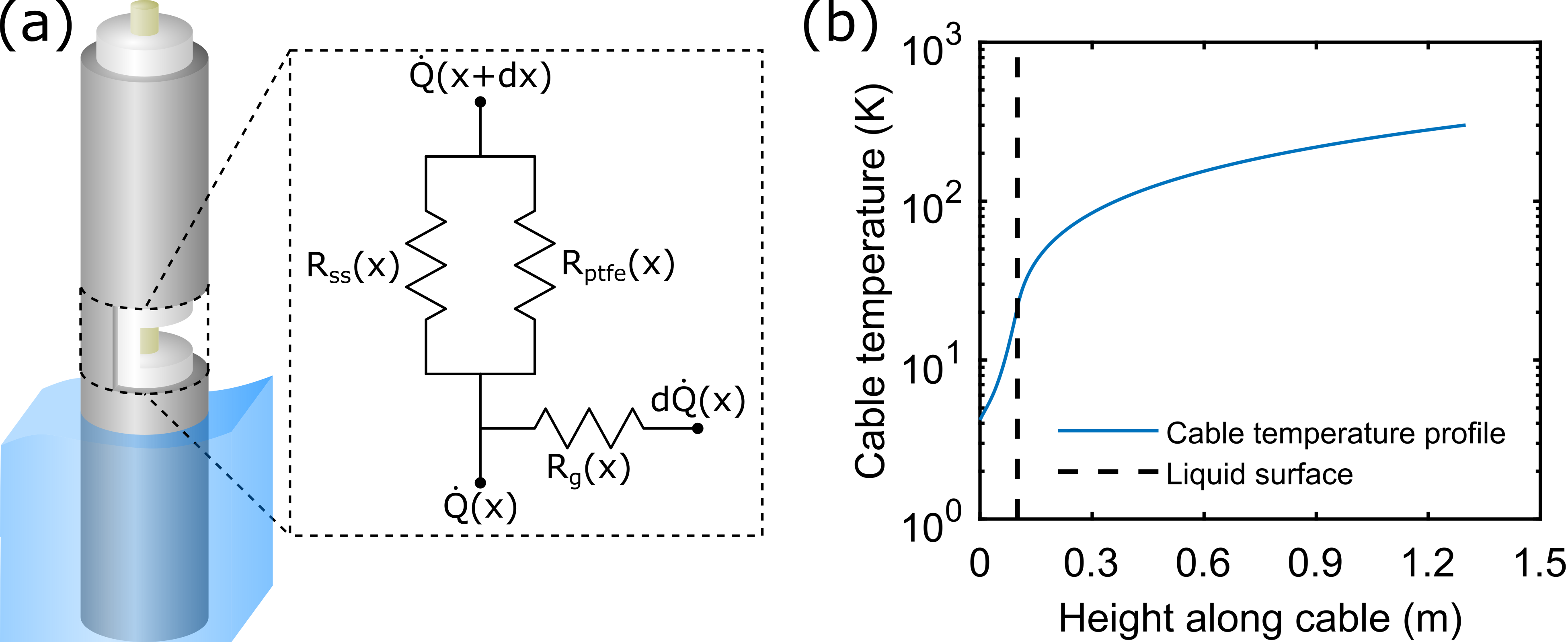}
    {\phantomsubcaption\label{fig:TempProfSchem}}
    {\phantomsubcaption\label{fig:TempProfPlot}}
    \caption{\textbf{(a)} Schematic of the coaxial cable temperature model showing a slice of length dx along the cable in contact with the gas environment. \textbf{(b)} Coaxial cable temperature profile (blue line) versus height along the cable. The height of the liquid surface is also shown (vertical dashed black line).}
\end{figure}

A schematic of the model is shown in \cref{fig:TempProfSchem}, where we have defined $R_\text{steel}=(\kappa_\text{steel}A_\text{steel})^{-1}$, $R_\text{ptfe}=(\kappa_\text{ptfe}A_\text{ptfe})^{-1}$, and $R_\text{g,l}=(PH_\text{g,l}\text{d}x)^{-1}$. \Cref{eq:ThermProf} was solved numerically using the bvp4c routine in Matlab with fixed boundary values of 301 K and 4.2 K and assuming a liquid surface height of 10 cm above the bottom of the cable. \Cref{tab:thermparams} lists the remaining assumed parameter values, and \cref{fig:TempProfPlot} shows the modelled coaxial cable temperature profile. Averaging over this curve yields an effective lumped physical temperature of the stainless steel coaxial cable of $T_\text{steel}=208$ K.

We now determine the effective lumped physical temperature of the full cable by including the additional SMA cabling which connected the stainless steel cables to the noise source and backend. We make the simplifying assumptions that the entire stainless steel cable radiates its noise at $T_\text{steel}$ with a loss $L_\text{steel}=2.34$ dB and that the additional cabling radiates entirely at $T_\text{cab}=301$ K with a loss of $L_\text{cab}=0.65$ dB. The additional cabling loss was measured directly, and the stainless steel cable loss was found by subtracting $L_\text{cab}$ from the total loss measured at 4.2 K shown in Fig. \textcolor{cyan}{2(a)}, at 5 GHz. Cascading the noise from these two cables and applying \cref{eq:Tatten}, we find: 
\begin{equation}
    \begin{gathered}
        T_\text{coax} = \frac{L_\text{cab}L_\text{steel}}{L_\text{cab}L_\text{steel}-1}\bigg(\frac{T_\text{cab}(L_\text{cab}-1)}{L_\text{cab}L_\text{steel}} + \frac{T_\text{steel}(L_\text{steel}-1)}{L_\text{steel}}\bigg)
    \end{gathered}
\end{equation}
yielding $T_\text{coax} = 223.3$ K, which was used for the value of the horizontal line shown in Fig. \textcolor{cyan}{2(b)}. 

\begin{center}
\begin{table}
\begin{tabular}{M{0.25\textwidth} M{0.25\textwidth}}
\hline
\hline
\text{Parameter} & \text{Value} \\
\hline
$\kappa_\text{steel}$ & $T$ dependent \cite{noauthor_nist_nodate-1}\\
$\kappa_\text{ptfe}$ & $T$ dependent  \cite{noauthor_nist_nodate}\\
$A_\text{steel}$ & 1.30 mm$^2$\\
$A_\text{ptfe}$ & 18.83 mm$^2$\\
$P$ & 1.12 mm\\
$H_\text{g}$ & 30 Wm$^{-2}$K$^{-1}$ \cite{jiji_heat_2009}\\
$H_\text{l}$ & 15 kWm$^{-2}$K$^{-1}$ \cite{van_sciver_helium_2012}\\
\hline
\hline
\end{tabular}
\captionsetup{justification=centering}
\caption{Table of parameters used in solving the coaxial cable thermal model. \label{tab:thermparams}}
\end{table}
\end{center}

\clearpage
\section{Error analysis}\label{sec:error}
Here we estimate the contribution of each quantity in Eq. (\textcolor{cyan}{3}) from the main text to the overall noise temperature measurement uncertainty $\Delta T_e$, as given by the standard error propagation formula: 
\begin{equation} \label{eq:LinErr}
\Delta T_e^X = \bigg|\frac{\partial T_e}{\partial X}\bigg| \Delta X
\end{equation}
where X denotes the measurement error source. All individual error sources are assumed to be independent unless otherwise stated, and they are added in quadrature to estimate the overall uncertainty. Numerical estimates listed below assume a noise temperature of $T_e=2$ K. We assume that each VNA loss measurement has an uncertainty of $\pm 0.01$ dB, which is the magnitude of the variation in the measured loss versus frequency of the calibration cable immediately after calibration. We also assume that the variance in the measured losses across different calibration and measurement dewars is 0.03 dB.

\subsection{Attenuator}
The uncertainty in the attenuator loss $\Delta L_2$ is found by adding the uncertainty from the VNA measurements in calibration dewars 1 and 2 in quadrature, so that $\Delta L_2 = 0.05$ dB.
\begin{gather}\label{eq:L2_err}
    \bigg|\frac{\partial T_e}{\partial L_2}\bigg| = \frac{T_e + T_{L_2}}{L_2}
\end{gather} 
yielding $\Delta T_e^{L_2} = 0.036$ K.

The uncertainty in the attenuator temperature $T_{L_2}$ is determined by the temperature diode calibration. The calibrated Lake Shore DT-670-SD bonded to the attenuator chip has a manufacturer-reported temperature uncertainty of $\Delta T_{L_2}=\pm20$ mK.
\begin{gather} \label{eq:T_L2_err}
\bigg|\frac{\partial T_e}{\partial T_{L_2}}\bigg| = \frac{L_2-1}{L_2}
\end{gather}
yielding $\Delta T_e^{T_{L_2}} = 0.020$ K.

\subsection{Coaxial Cables}
The uncertainties in the coaxial cable losses $\Delta L_1$ and $\Delta L_3$ are found by adding the uncertainties from three separate VNA loss measurements: the measurement of $L_\text{coax}=L_1L_3$ in calibration dewar 1, and the measurements of $L_1$ and $L_3$ at room temperature to determine the loss asymmetry between the two cables, yielding $\Delta L_1 = 0.06$ dB and $\Delta L_3 = 0.07$ dB.
\begin{gather} \label{eq:L1_err}
\bigg|\frac{\partial T_e}{\partial L_1}\bigg| = \frac{T_e + T_{L_1}L_2^{-1} + T_{L_2}(L_2-1)L_2^{-1}}{L_1}
\end{gather}
\begin{gather} \label{eq:L3_part}
\bigg|\frac{\partial T_e}{\partial L_3}\bigg| = \frac{T_{L_3}}{G_\text{full}L_1L_2L_3^2}
\end{gather}
yielding $\Delta T_e^{L_1} = 0.088$ K and $\Delta T_e^{L_3} = 0.006$ K.

The uncertainty in the cable temperatures $\Delta T_{L_1}$ and $\Delta T_{L_3}$ are derived from error analysis of \cref{eq:TL}. We estimate $\Delta T_\text{coax}=\pm20$ K.
\begin{gather} \label{eq:T_L1_err}
\bigg|\frac{\partial T_e}{\partial T_{L_1}}\bigg| = \frac{L_1-1}{L_1L_2}
\end{gather}
\begin{gather} \label{eq:T_L3_err}
\bigg|\frac{\partial T_e}{\partial T_{L_3}}\bigg| = \frac{L_3-1}{G_\text{full}L_1L_2L_3}
\end{gather}
yielding $\Delta T_e^{T_{L_1}} = 0.100$ K and $\Delta T_e^{T_{L_3}} = 0.032$ K.

\subsection{Gain}
The uncertainty in the total gain $\Delta G_\text{full}$ comes directly from the uncertainty of a single VNA loss measurement so that $\Delta G_\text{full}=\pm0.01$ dB.
\begin{gather} \label{eq:Gfull_err}
\bigg|\frac{\partial T_e}{\partial G_\text{full}}\bigg| = \frac{T_\text{coax}(L_3-1)L_3^{-1} + T_\text{BE}}{L_1L_2G_\text{full}^2}
\end{gather}
yielding $\Delta T_e^{G_\text{full}} = 0.002$ K.

\subsection{Back-end detector}
The uncertainty in the backend detector noise temperature $\Delta T_\text{BE}$ is determined by the temperature and loss uncertainties in the coaxial cable connecting the cooled load to the backend. We estimate $\Delta T_\text{BE}=\pm5$ K.
\begin{gather} \label{eq:T_BE_err}
\bigg|\frac{\partial T_e}{\partial T_{\text{BE}}}\bigg| = \frac{1}{G_\text{full}L_1L_2}
\end{gather}
yielding $\Delta T_e^{\text{BE}} = 0.016$ K.

\subsection{Noise source}
The uncertainty in the noise source ENR comes from error analysis of \cref{eq:ENR}. We estimate $\Delta E=\pm0.040$ dB.
\begin{gather} \label{eq:ENR_err}
\bigg|\frac{\partial T_e}{\partial E}\bigg| = \frac{T_0}{(Y-1)L_1L_2}
\end{gather}
yielding $\Delta T_e^{E} = 0.073$ K.

The uncertainty in the noise source diode temperature $\Delta T_\text{C}$ comes from the uncertainty in the Type T thermocouple temperature measurement of the noise source chassis. We estimate $\Delta T_\text{C}=1$ K.
\begin{gather} \label{eq:TC_err}
\bigg|\frac{\partial T_e}{\partial T_\text{C}}\bigg| = \frac{1}{L_1L_2}
\end{gather}
yielding $\Delta T_e^{T_C} = 0.005$ K.

\subsection{Y-Factor power}
The Y-factor measurement uncertainty $\Delta Y$ accounts for all uncertainty sources originating after the transduction of microwave power to DC voltage. We report an effective normalized Y-factor error $\frac{\Delta Y}{Y}$ of better than $3\times 10^{-4}$ for a 4 s integration time, which used for all steady-state data presented in this paper.
\begin{gather} \label{eq:Y_err}
\bigg|\frac{\partial T_e}{\partial Y}\bigg| = \frac{T_0E}{L_1L_2(Y-1)^2}
\end{gather}
yielding $\Delta T_e^{\text{Y}} = 0.003$ K.

\subsection{Cable mismatch}
There is error introduced from the difference in noise source impedance between the on and off state, which causes a changing reflection coefficient between the noise source and the first component in the measurement chain (in our experiment this is the input coaxial cable). In cases where the impedance match at the output of the noise source is poor, this error must be considered, and it can be corrected for if the full S-parameters of the noise source in the on and off state and of the cable are known. In our experiment this error was found to contribute negligibly to the overall uncertainty.

\subsection{Overall uncertainty}
The uncertainty budget is shown in \cref{tab:errors}. The uncertainty analysis shown in this section was used to generate the error bars in the primary noise temperature datasets.

\begin{center}
\begin{table}[!b]
\begin{tabular}{M{0.24\textwidth} M{0.24\textwidth} M{0.24\textwidth} M{0.24\textwidth}}
\hline
\hline
\text{Error source} & \text{Value} & \text{Estimated error} & \text{Contribution to $T_e$} \\
\hline
$L_2$ & $20.00$ dB & $0.05$ dB & $0.036$ K\\
$T_2$ & $1.600$ K &  $0.020$ K & $0.020$ K\\
$L_1$ & $3.25$ dB &  $0.06$ dB & $0.088$ K\\
$L_3$ & $3.44$ dB &  $0.07$ dB & $0.006$ K\\
$T_1$ & $223$ K &  $20$ K & $0.100$ K\\
$T_3$ & $223$ K &  $20$ K & $0.032$ K\\
$G_\text{full}$ & $1.98$ dB &  $0.01$ dB & $0.002$ K\\
$T_\text{BE}$ & $170$ K &  $5.0$ K & $0.016$ K\\
$E$ & $15.0$ dB &  $0.040$ dB & $0.073$ K\\
$T_\text{C}$ & $301.0$ K &  $1$ K & $0.009$ K\\
$Y$ & $6.8$ &  $0.002$ &  $0.003$ K\\
RMS Sum & & & $0.162$ K\\
\hline
\hline
\end{tabular}
\captionsetup{justification=centering}
\caption{Table of parameters used to extract $T_e$, and their associated uncertainties. \label{tab:errors}}
\end{table}
\end{center}

\clearpage

\bibliographystyle{aip}
\bibliography{references}